\documentclass[twocolumn]{IEEEtran}
\usepackage{cite}
\usepackage[cmex10]{amsmath}
\usepackage{array}
\usepackage{amssymb}
\usepackage{amsfonts}
\usepackage{subeqnarray}
\usepackage{cases}
\usepackage{mathrsfs}
\usepackage{graphicx}
\usepackage{threeparttable}
\usepackage{acronym}
\usepackage{caption}
\usepackage{subcaption}
\usepackage{dsfont}
\usepackage[linesnumbered,ruled,vlined]{algorithm2e}
\begin{document}

\title{Indoor occupancy estimation from carbon dioxide concentration}

\author{\IEEEauthorblockN{Chaoyang Jiang, Mustafa K. Masood, Yeng Chai Soh, and Hua Li}

\thanks{The work was supported by Singapore's National Research Foundation under NRF2011NRF-CRP001-090, and partially supported by the Energy Research Institute at NTU(ERI@N).

C. Jiang, M. K. Masood and Y. C. Soh are with the School of Electrical and Electronic Engineering, Nanyang Technological University, 639798 Singapore (e-mail: chaoyangjiang@hotmail.com; must0006@e.ntu.edu.sg; eycsoh@ntu.edu.sg).

H. Li is with the school of Mechanical and Aerospace Engineering, Nanyang Technological University, 639798 Singapore (e-mail: lihua@ntu.edu.sg.).}
}

\maketitle

\begin{abstract}
This paper presents an indoor occupancy estimator with which we can estimate the number of real-time indoor occupants based on the carbon dioxide (CO2) measurement. The estimator is actually a dynamic model of the occupancy level. To identify the dynamic model, we propose the Feature Scaled Extreme Learning Machine (FS-ELM) algorithm, which is a variation of the standard Extreme Learning Machine (ELM) but is shown to perform better for the occupancy estimation problem. The measured CO2 concentration suffers from serious spikes. We find that pre-smoothing the CO2 data can greatly improve the estimation accuracy. In real applications, however, we cannot obtain the real-time globally smoothed CO2 data. We provide a way to use the locally smoothed CO2 data instead, which is real-time available.  We introduce a new criterion, i.e. $x$-tolerance accuracy, to assess the occupancy estimator. The proposed occupancy estimator was tested in an office room with 24 cubicles and 11 open seats. The accuracy is up to 94 percent with a tolerance of
 four occupants.
\end{abstract}

\IEEEpeerreviewmaketitle

\section{Introduction}

\IEEEPARstart{I}{ndoor} occupancy information is an important part of home and office automation. It can be used as an input for the control of indoor lighting systems \cite{candanedo2016accurate, nguyen2013energy} and Heat, Ventilation and Air-conditioning (HVAC) systems \cite{ebadat2015regularized, oldewurtel2013importance}. Studies have shown that around one-third of the energy consumed in buildings can be saved using occupancy-based control \cite{brooks2015energy, brooks2014experimental, erickson2011observe, dong2009sensor, erickson2014occupancy}. Therefore, indoor occupancy estimation has become a popular area of research in recent years.

With multi-camera video and pattern recognition techniques, the number of indoor occupants can be accurately estimated \cite{fleuret2008multicamera, liu2013measuring, benezeth2011towards, tomastik2008video}. But these type of methods are intrusive and cannot be widely used due to the expensive hardware cost and privacy concerns. A number of terminal-based methods have also been proposed in which occupants are required to use a device, such as the keyboard and mouse \cite{nguyen2012beyond, zhao2015virtual}, smartphones with WIFI \cite{depatla2015occupancy} or Radio Frequency Identification (RFID) tags \cite{agarwal2011duty, martani2012enernet, li2012measuring}. With these methods, data security can be an issue. Also, they cannot detect occupants not using the devices. Consequently, it is imperative to find a non-intrusive and non-terminal-based method for indoor occupancy estimation.

This paper provides a new indoor occupancy estimator based on real-time CO2 measurement. CO2 sensors are available for standard HVAC systems and hence no additional hardware cost is required. More importantly, CO2 sensors are non-intrusive and non-terminal-based. Next, we briefly review the current non-intrusive and non-terminal-based occupancy estimators.

\subsection{Related prior work}

Pyroelectric infrared (PIR) sensors are often used for human detection. They can detect the infrared radiation emitted from human beings within their field of view. When the sensor view is impeded PIR sensors are prone to false estimates \cite{teixeira2010survey}. If many occupants are together, PIR sensors cannot tell the exact number \cite{teixeira2010survey, guo2010performance}. Therefore, with PIR sensors it is easy to determine whether a room is occupied \cite{liuoccupancy} but difficult to estimate the number of occupants \cite{li2012measuring, teixeira2010survey, guo2010performance, labeodan2015occupancy}. PIR sensors usually have to be combined with other types of sensors for occupancy estimation \cite{agarwal2011duty, hailemariam2011real, yang2014systematic}.

Ultrasonic sensors like sonar are commonly used for motion detection. They emit ultrasonic sound waves and receive the reflected signals. When an object moves, the wavelength of the reflected signal would be different from the emitted signal. This technique was used for indoor occupancy detection \cite{ul2014review, tarzia2009sonar} but cannot determine the number of occupants \cite{labeodan2015occupancy}.

In \cite{uziel2013networked}, the authors estimated the number of indoor occupants using 8 microphones, which detected the sound wave of occupants. This method cannot detect silent occupants \cite{labeodan2015occupancy}.

The number of indoor occupants can be predicted using the Agent Based Model (ABM) \cite{liao2011novel,liao2012agent} and inhomogeneous Markov chain model \cite{page2008generalised, chen2015modeling} based on the statistical information of the historical occupancy data. These methods require no real-time sensor data but a large amount of training data.

Recently, estimating the number of indoor occupants from environmental parameters such as temperature, humidity, pressure and CO2 concentration is becoming popular. Environmental parameter data can be directly obtained from HVAC systems. Also, it is non-intrusive and non-terminal-based. In \cite{candanedo2016accurate, dong2010information, lam2009occupancy, ekwevugbe2013real, masood2015real}, features (such as first order difference and second order difference) were extracted from environmental parameters and machine learning technologies (such as support vector machine (SVM), artificial neural networks (ANN) and hidden Markov models (HMM)) were used to construct the relation between the features and the occupancy level. The number of occupants was considered to be depending on the current state and features of environmental parameters. This approach is effective in detecting the presence of occupants \cite{candanedo2016accurate} but is inaccurate in estimation of numbers, even when the maximum number of the occupants is small (the accuracy is less than 75$\%$ for a maximum of 4 occupants in \cite{dong2010information, lam2009occupancy}).

A recent work identified a linear dynamic model of the indoor temperature and CO2 concentration in which the occupancy level is an input \cite{ebadat2015regularized}. With such a dynamic model, the number of occupants can be estimated by solving a deconvolution problem. This method was tested in a room with a maximum of 4 occupants. The occupancy level were estimated with 5 minutes time delay and the accuracy was up to 88.8$\%$. In this method, both the model identification and the deconvolution processes can only provide approximate results, which motivates us to directly identify an occupancy estimator.

\subsection{Statements of our contributions}
In this paper, we present an estimator which can tell us the number of real-time indoor occupants. The estimator is actually a dynamic model of the occupancy level in which the current occupant number depends on the measurements of CO2 concentration, and the estimated occupancy level in a past time horizon. In this way, no deconvolution process is required.

We have done an experiment in a $9.3m\times 20m$ air-conditioned office room with 24 office cubic and 11 open seats. The experimental results verify the effectiveness of the proposed estimator. To the best of our knowledge, most of the current work deals  with very few occupants \cite{ebadat2015regularized, dong2010information, lam2009occupancy}, and no non-intrusive and non-terminal-based method has been shown to be effective for rooms with more than 20 occupants.

To identify the occupancy estimator, we modify the standard Extreme Learning Machine (ELM) by adding a feature layer. The feature-to-hidden layer weight matrix is designed as a scaled random matrix.
The modified ELM, called \emph{feature scaled ELM} (FS-ELM), retains the computational efficiency of the standard ELM and is shown to be more effective to identify the occupancy estimator. The application of the FS-ELM can be easily extended to other problems.

The measured CO2 concentration sometimes suffers from serious spikes, which has a negative influence on the occupancy estimation. We found that smoothing the CO2 data before the training process can greatly improve the performance of the occupancy estimator.  To smooth the CO2 data, the global information of CO2 concentration in the time domain is required. However, when using the estimator, the real-time globally smoothed CO2 data is unavailable because the future measurements is unknown. We used the locally smoothed CO2 data to replace the globally smoothed data and provided one way to remove the accumulated error deduced by the replacement.

In addition, we introduce the notion of $x$-\emph{tolerance accuracy} to assess the results of the occupancy estimator.

\subsection{Organization of the paper}
The rest of this paper is organized as follows. In Section II, we introduce the problem of indoor occupancy estimation from CO2 measurements. In Section III, we present the FS-ELM with which the proposed occupancy estimator is identified. In Section IV, we identify the occupancy estimator from globally smoothed CO2 data, and provide a way to use the estimator based on the real-time locally smoothed CO2 data.  We then show the experiment verification in Section V, and the conclusions are given in Section VI.

\section{Indoor Occupancy estimation from CO2 data}

Various dynamic models of the indoor CO2 concentration have been discussed in the literature \cite{ebadat2015regularized, ansanay2013estimating, gruber2014co, scotton2013physics, cali2015co, kar2011accurate}. In summary, we introduce the following generalized discrete-time state space model
\begin{equation}\label{eq:CO2DModel}
  c_k = g(\mathbf{c}_{k-l:k-1}, \mathbf{o}_{k-l:k}, \mathbf{v}_{k-l:k})
\end{equation}
where $c_k$ is the CO2 concentration around an indoor sensor node at time instant $t_k$, $g(\cdot)$ is a unknown function, and $\mathbf{c}_{k-l:k-1} =[c_{k-l}, c_{k-l+1},\cdots, c_{k-1}]^\mathrm{T}$ is the sequence of CO2 concentration at the past time horizon $[t_{k-l}, t_{k-1}]$. Similarly, $\mathbf{o}_{k-l:k}=[o_{k-l}, o_{k-l+1},\cdots,o_{k}]^\mathrm{T}$ and $\mathbf{v}_{k-l:k}=[v_{k-l}, v_{k-l+1},\cdots, v_{k}]^\mathrm{T}$ are the sequence of indoor occupants number and venting level, respectively. In the standard HVAC system, the ventilation system is controlled based on the measured CO2 concentration. Hence, the venting level can be estimated from the CO2 measurements. In addition, there is another type of ventilation system which delivers a constant supply of fresh air, and the venting level is known \emph{a priori}.

In \cite{ebadat2015regularized, ansanay2013estimating, gruber2014co, scotton2013physics, cali2015co, kar2011accurate} the indoor CO2 concentration is assumed to be uniform, and in \cite{ansanay2013estimating, gruber2014co, scotton2013physics, cali2015co} the length of time horizon $l=1$, which implies that indoor CO2 concentration is with Markov property (memoryless property ).
However, \cite{weekly2015modeling} showed that the gradient of indoor CO2 concentration can be very large, and CO2 emitted by certain occupant cannot be immediately sensed. Therefore, it is unreasonable to simply set $l=1$ unless the sampling time is large enough.

Apparently, if the CO2 dynamic model has been identified, the number of indoor occupants can be estimated based on the real-time CO2 measurements. In \cite{ebadat2015regularized}, the estimator is designed by solving a deconvolution problem. In this strategy, the deconvolution process suffers from truncation errors and the error of the identified CO2 dynamic model. Therefore, we intend to directly identify the occupancy estimator, thus  avoiding the deconvolution process.

Considering the CO2 dynamic model \eqref{eq:CO2DModel}, one generalized occupancy estimator can be described as
\begin{equation}\label{eq:estimator}
  o_k = f(\mathbf{c}_{k-l:k}, \mathbf{o}_{k-l:k-1}, \mathbf{v}_{k-l:k}) = f(\mathbf{x}_k)
\end{equation}
where $f(\cdot)$ is the model to be identified, and
\begin{equation}
  \mathbf{x}_{k}=[\mathbf{c}_{k-l:k}^\mathrm{T}\;\; \mathbf{o}_{k-l:k-1}^\mathrm{T}\;\; \mathbf{v}_{k-l:k}^\mathrm{T}]^\mathrm{T}\in\mathbb{ R}^n \label{eq:X_input}
\end{equation}
is the input of the occupancy estimator. Here $n=3l+2$ is the dimension of the input vector. The occupancy estimator is a regression model between $\mathbf{x}_k$ and the number of current occupants $o_k$.

The main work of the estimator design is to identify the model $f(\cdot)$, which is a regression problem.
With enough training samples, which can be off-line measured, many machine learning techniques like ANN \cite{ebadat2015regularized, dong2010information, lam2009occupancy, ekwevugbe2013real}, SVM \cite{ebadat2015regularized, dong2010information, lam2009occupancy}, ELM \cite{masood2015real}, and deep learning \cite{lecun2015deep} can be used to solve this regression problem.  In this work, we select the ELM due to its simplicity, computational efficiency, and flexibility. Next, we present a \emph{feature scaled extreme learning machine} (FS-ELM) algorithm, which is a variation of the standard ELM, for the model identification.

\section{FS-ELM for occupancy estimator identification}\label{Sec:Identification}

\subsection{The standard ELM for occupancy estimation}

The standard ELM is a fast way to train the single-hidden-layer feedforward networks (SLFNs). When using ELM the input-to-hidden layer connection weights are randomly generated without tuning and are independent of the training data. It is the reason why ELM is computationally much more efficient than many other machine learning techniques in the training process. This idea has been discussed in \cite{schmidt1992feedforward, pao1992functional, pao1994learning, igelnik1995stochastic}, and ELM was first coined in \cite{huang2004extreme}. For more details about ELM, do refer to \cite{huang2004extreme, huang2006extreme, huang2014insight, huang2015trends}.

\begin{figure}[ht]
    \centering
    \includegraphics[width=0.38\textwidth]{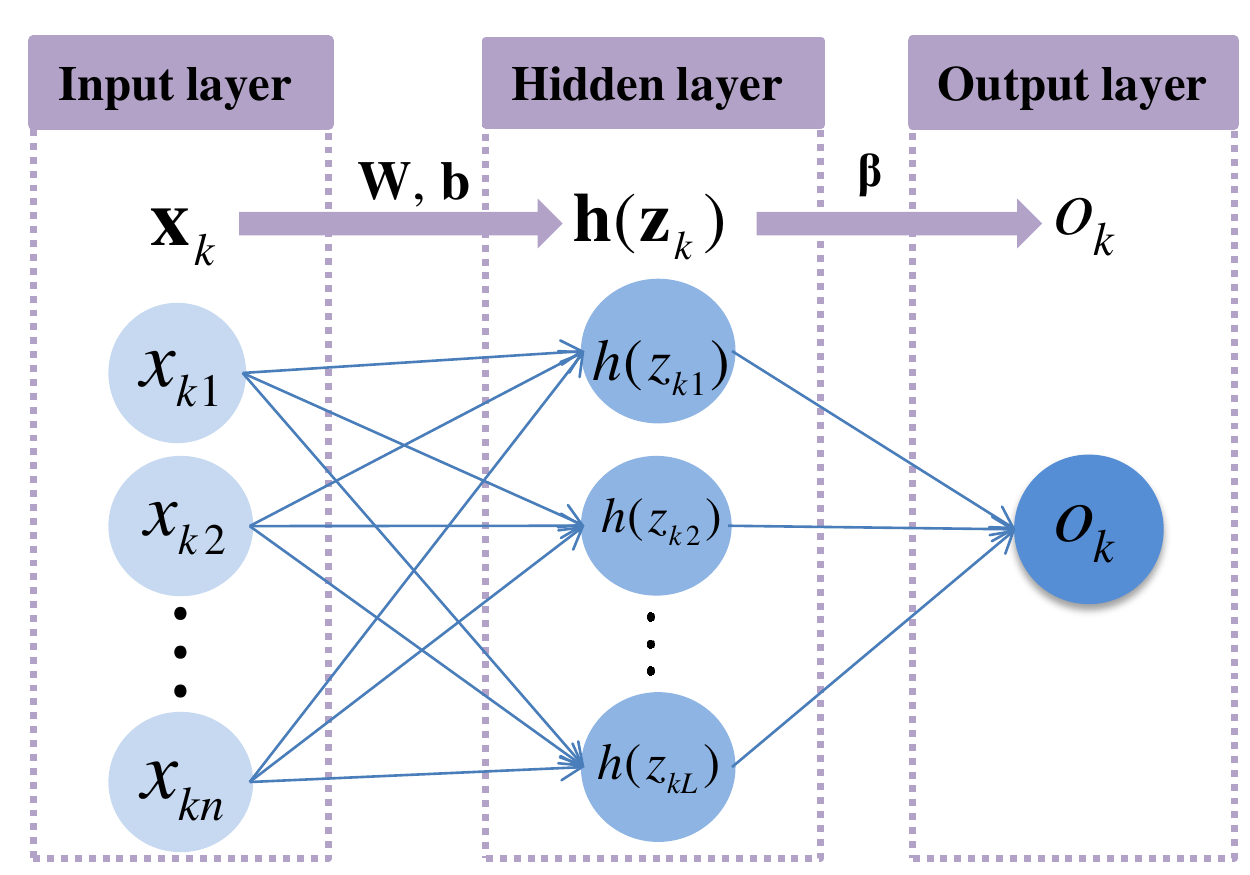}
       \caption{The structure of the  standard extreme learning machine (ELM).}
    \label{Fig:ELM}
\end{figure}

When using the standard ELM, the occupancy estimator is a SLFN as shown in Fig. \ref{Fig:ELM}.
The output of the SLFN, i.e. the number of occupants, can be formulated as
\begin{equation}\label{eq:ELMoutput}
  o_k = f(\mathbf{x}_k)= \sum_{i=1}^L \beta_ih(\mathbf{w}_i^\mathrm{T}\mathbf{x}_k+b_i)
\end{equation}
where $L$ is the number of hidden neurons, $\mathbf{w}_i\in\mathbb{R}^n$ consists of the random weights from the input layer to the $i$-th hidden neuron, $b_i$ is the random bias of the input of $i$-th hidden neuron, $h(\cdot)$ is a known activation function called  \emph{ELM random feature mapping} \cite{huang2014insight}, and $\boldsymbol \beta = [\beta _1, \beta_2,\cdots,\beta_L]^\mathrm{T}$ represents the hidden-to-output connection weights. With the training data set, $\boldsymbol\beta$ can be easily found by solving a least-squares problem.

\begin{figure}[ht]
    \centering
    \includegraphics[width=0.40\textwidth]{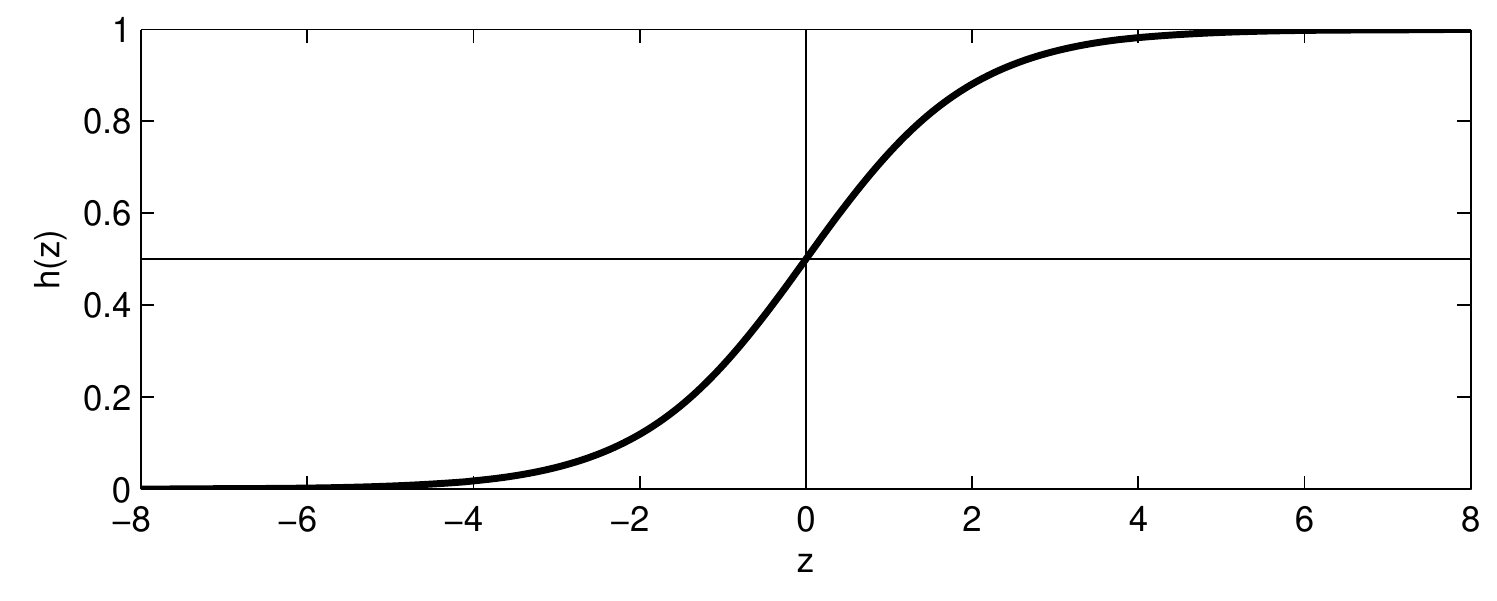}
       \caption{The sigmoid function.}
    \label{Fig:SigmoidFun}
\end{figure}

As shown in Fig. \ref{Fig:ELM}, $z_{ki}$ is the input of the activation function, and $z_{ki}=\mathbf{w}_i^\mathrm{T}\mathbf{x}_k+b_i$. A commonly used  activation function is the sigmoid function
\begin{equation*}
  h(z)=\frac{1}{1+\mathrm{exp}(-z)}
\end{equation*}
which is shown in Fig. \ref{Fig:SigmoidFun}. We can find from Fig. \ref{Fig:SigmoidFun} that $h(z_{ki})\approx 0$ and $h(z_{ki})\approx 1$ for all $z_{ki}<-5$ and $z_{ki}>5$, respectively. Therefore, for any $\mathbf{x}_k$ which can guarantee $|z_{ki}|>5$, the output of the hidden neuron will be 0 or 1, and the result is non-sensitive to the exact value of $\mathbf{x}_k$. In other words, ELM cannot distinguish the input layer data $\mathbf{x}_k$ if $\mathbf{x}_k$ can lead to $|z_{ki}|>5$.

For some occupancy estimation problems \cite{ebadat2015regularized, masood2015real}, when using standard ELM to train the data set, we found that most of the hidden neurons of the SLFNs are always 0 or 1, which leads to a low accuracy estimation. Considering the ELM universal approximation capability theorems \cite{huang2006universal, huang2007convex, huang2008enhanced, huang2014insight}, if we use enough hidden neurons we may still obtain accurate results. However, with more hidden neurons, the computational cost in the training process is higher. This phenomenon may have no negative influence on a simple classification problem, but it does lead to poor results for many regression problems. All the other  \emph{ELM random feature mappings} mentioned in \cite{huang2014insight} suffer from similar problems. In this work, we only consider the sigmoid function, and similar steps can be taken if other activation functions are used.

Let $\mathbf{z}_k=[z_{k1}, z_{k2},\dots,z_{kL}]^\mathrm{T}$, it is clear that
\begin{equation}\label{eq:zk}
  \mathbf{z}_k = \mathbf{W}\mathbf{x}_k+\mathbf{b}
\end{equation}
where $\mathbf{W}=[\mathbf{w}_1\;\mathbf{w}_2\dots \mathbf{w}_L]^\mathrm{T}$ and $\mathbf{b}=[b_1\; b_2\dots b_L]^\mathrm{T}$. To make the hidden neurons more meaningful, i.e. to make sure that the output of hidden neurons can reflect the diversity of input layer data, we should guarantee that
\begin{equation}\label{eq:requirement}
  \|\mathbf{z}_k\|_{\infty}<5
\end{equation}
This requirement implies that the input-to-hidden layer connection weights are dependent on the training set, and thus they cannot be randomly set.

In addition, the input layer data $\mathbf{x}_k$ consists of three types of data, and the magnitude of each type of data may differ significantly from others. The indoor CO2 concentration is typically from several hundred to a few thousand ppm (parts-per-million, i.e. the unit of CO2 concentration), while the values of both the occupancy level and venting level are much smaller. In such a case, if the input-to-hidden weight matrix is randomly generated the influence of CO2 concentration will overwhelm that of occupancy levels and venting levels, which would adversely affect the training of the regression model.

To improve the performance and retain the computational efficiency, we introduce two steps to modify the standard ELM: 1) add a feature layer between the input layer and the hidden layer, and 2) scale the randomly generated feature-to-hidden layer connection weights. We termed the modified ELM as \emph{feature scaled extreme learning machine} (FS-ELM). The structure of the proposed FS-ELM is shown in Fig. \ref{Fig:SELM}, and the details of the FS-ELM are shown below.

\begin{figure}[ht]
    \centering
    \includegraphics[width=0.48\textwidth]{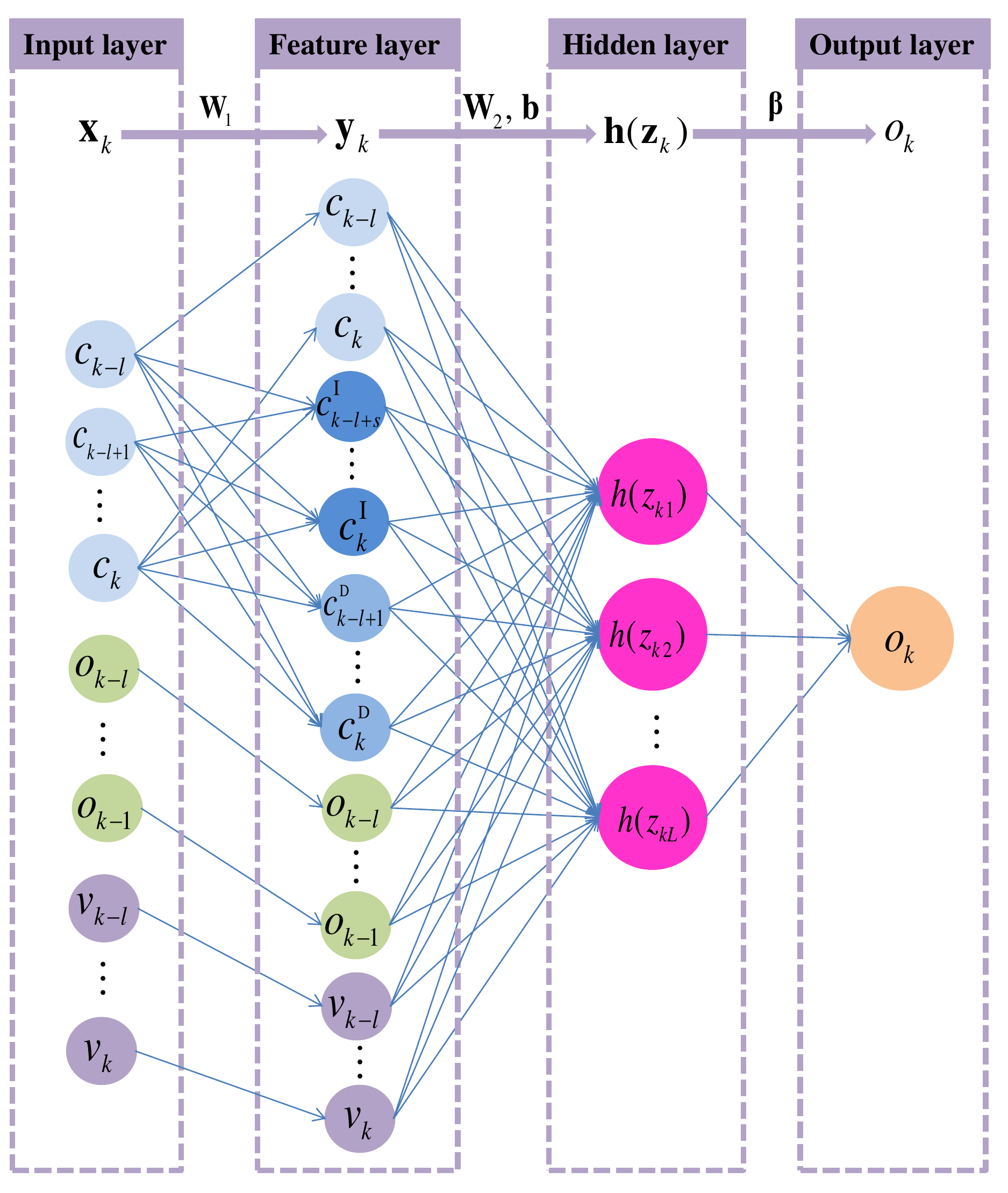}
       \caption{The structure of the  feature scaled extreme learning machine (FS-ELM).}
    \label{Fig:SELM}
\end{figure}

\subsection{The feature layer of the FS-ELM}
Both ELM and other machine learning techniques like ANN and SVM are not `intelligent' enough to accurately estimate the real-time occupancy level directly from the raw measured environmental parameters (CO2 concentration, temperature, and relative humidity, etc.). But from some properly selected features of the environmental parameters, these machine learning techniques can provide better results \cite{dong2010information, lam2009occupancy, ekwevugbe2013real, masood2015real}.
All the features can be obtained from the input layer data, i.e. $\mathbf{x}_k$.

For general regression problems, the way to automatically find the best features is still an open problem in the machine learning community.
In this work, we add a feature layer in between the input-hidden layers for a SLFN. Based on the \emph{a priori} discussions for the occupancy estimation \cite{dong2010information, lam2009occupancy, ekwevugbe2013real, masood2015real}, we summarize the features of CO2 concentration as follows.
\begin{itemize}
    \item \underline{CO2 data:} the sequence of CO2 concentration in a past time horizon, i.e. $\mathbf{c}_{k-l:k}$.
    \item \underline{Integration of the CO2 differential:} the sequence of integration of the first-order difference of CO2 concentration in a time interval, i.e. $\mathbf{c}_{k-l+s:k}^\mathrm{I}=\mathbf{c}_{k-l+s:k}-\mathbf{c}_{k-l:k-s}$.
    \item \underline{Difference of CO2 data:} the sequence of first-order difference of CO2 data, i.e. $\mathbf{c}_{k-l+1:k}^{\mathrm{D}}=\mathbf{c}_{k-l+1:k}-\mathbf{c}_{k-l:k-1}$.
\end{itemize}
We call the three features as `PID of CO2 data'.
Actually both $\mathbf{c}_{k-l+s:k}^\mathrm{I}$ and $\mathbf{c}_{k-l+1:k}^{\mathrm{D}}$ are the difference of CO2 data in the time domain, but the former is with a larger sampling time ($s$). Therefore, we can view the entries of $\mathbf{c}_{k-l+s:k}^\mathrm{I}$ as the integration of the differential of CO2 data in a time interval with length $s$.

The PID of CO2 data together with the past occupancy level and the venting level constitute the feature layer, as shown in the second layer in Fig. \ref{Fig:SELM}. For simplicity, we denote the feature layer data as the vector
\begin{eqnarray}
\nonumber \mathbf{y}_k = \left[\mathbf{c}_{k-l:k}^\mathrm{T}\;\;\;\; (\mathbf{c}_{k-l+s:k}^\mathrm{I})^\mathrm{T} \;\;\;\; (\mathbf{c}_{k-l+1:k}^{\mathrm{D}})^\mathrm{T} \quad \quad \quad \quad  \right.\\ \left.\mathbf{o}_{k-l:k-1}^\mathrm{T}\quad \mathbf{v}_{k-l:k}^\mathrm{T}\right]^\mathrm{T} \label{eq:ycomponents}
\end{eqnarray}
where $\mathbf{y}_k\in \mathbb{R}^{n_\mathrm{f}}$ and $n_\mathrm{f}=5l-s+3$. From \eqref{eq:X_input} and \eqref{eq:ycomponents} we can find that
\begin{equation}\label{eq:featureLyer}
  \mathbf{y}_k=\mathbf{W}_1\mathbf{x}_k = \mathrm{blkdiag}\{\mathbf{W}_{\mathrm{PID}}, \mathbf{I}_{2l+1}\}\mathbf{x}_k
\end{equation}
where $\mathbf{I}_n$ is the $n$-dimensional identity matrix, $\mathrm{blkdiag}\{\cdot\}$ is the block diagonal matrix operator, and
\begin{equation*}\label{eq:PID}
  \mathbf{W}_{\mathrm{PID}}=\left[\mathbf{I}_{l+1}\quad
  [\!\begin{array}{c}
       \mathbf{0} \\
       \mathbf{I}_{l-s+1} \\
     \end{array}\!] \!-\!
   [\!\begin{array}{c}
                     \mathbf{I}_{l-s+1} \\
                     \mathbf{0} \\
                   \end{array}\!] \quad
   [\!\begin{array}{c}
       \mathbf{0} \\
       \mathbf{I}_{l} \\
     \end{array}
   \!]\!-\!
   [ \! \begin{array}{c}
                     \mathbf{I}_l \\
                     \mathbf{0} \\
                   \end{array}\!]\right]^\mathrm{T}
\end{equation*}
where $\mathbf{0}$ represents a proper dimensional matrix with all zero entries.

\subsection{The hidden layer of the FS-ELM}
We can view the feature layer as a preprocessing of the input layer data, and the last three layers in Fig. \ref{Fig:SELM} (i.e., the feature layer, the hidden layer and the output layer) is the same as the structure of the standard ELM. However, the feature-to-hidden layer connection weights of FS-ELM are not a random matrix. The input of the activation function is
\begin{equation}\label{eq:zk}
  \mathbf{z}_k = \mathbf{W}_2\mathbf{y}_k+\mathbf{b}=\mathbf{RS}\mathbf{y}_k +\mathbf{b}
\end{equation}
where $\mathbf{W}_{2}=\mathbf{RS}\in \mathbb{R}^{L\times n_\mathrm{f}}$ is the feature-to-hidden layer weight matrix, $\mathbf{R}\in \mathbb{R}^{L\times n_\mathrm{f}}$ is a random matrix, and $\mathbf{S}\in \mathbb{R}^{n_\mathrm{f}\times n_\mathrm{f}}$ is a diagonal matrix. Here, $\mathbf{S}$ is unknown \emph{a priori} and we can design it to scale $\mathbf{z}_k$ and guarantee the requirement of \eqref{eq:requirement}.

The output of the activation function, i.e. the output of hidden layer is
\begin{equation*}
  \mathbf{h}(\mathbf{z}_k)=[h(z_{k1})\;\;h(z_{k2})\;\cdots \;h(z_{kL})]
\end{equation*}
where the \emph{ELM random feature mapping} $h(\cdot)$ is the sigmoid function.

\subsection{FS-ELM training}\label{SubSec:training}

The output of FS-ELM can be written as
\begin{equation}\label{eq:IS-ELM}
  o_k = \mathbf{h}(\mathbf{z}_k)\boldsymbol\beta
  =\mathbf{h}(\mathbf{RS}\mathbf{W}_1\mathbf{x}_k+\mathbf{b})\boldsymbol\beta
\end{equation}
where $\mathbf{W}_1$ is known \emph{a priori}. Note that the entries of $\mathbf{b}$ and $\mathbf{R}$ are randomly generated. Therefore, the training process is to find the scale matrix $\mathbf{S}$ and the hidden-to-output layer weights $\boldsymbol\beta$.

Firstly, we determine the scale matrix.
To meet the requirement \eqref{eq:requirement}, we can simply set
\begin{equation}\label{eq:S_equal}
    \mathbf{S}=\frac{5}{ z_{\mathrm{max}}^\prime}\mathbf{I}_{n_\mathrm{f}}
\end{equation}
where $z_{\mathrm{max}}^\prime=\mathbf{max}\{\|\mathbf{z}_k^\prime\|_\infty, \; k\leq N\}$, $N$ is the number of training samples, and
\begin{equation*}
  \mathbf{z}_k^\prime = \mathbf{R}\mathbf{y}_k=\mathbf{RW}_1\mathbf{x}_k
\end{equation*}
Apparently, ignoring the bias vector $\mathbf{b}$, if $\mathbf{S}=\frac{5}{z_{\mathrm{max}}^\prime} \mathbf{I}_{n_\mathrm{f}}$ we can easily find that $\|\mathbf{z}_k\|_{\infty}\leq 5$ for all $k\leq N$. However, we can use the scale matrix $\mathbf{S}$ to adjust the influence of each feature. If such a diagonal matrix is carefully designed, we can obtain a better regression model.

We denote the scale matrix by
\begin{equation}\label{eq:S}
  \mathbf{S}=\mathrm{blkdiag}\{\alpha_\mathrm{P}\mathbf{I}_{l+1}, \alpha_\mathrm{I}\mathbf{I}_{l-s+1}, \alpha_\mathrm{D}\mathbf{I}_{l}, \alpha_\mathrm{o}\mathbf{I}_{l}, \alpha_\mathrm{v}\mathbf{I}_{l+1}\}
\end{equation}
where the parameters $\alpha_\mathrm{P}$, $\alpha_\mathrm{I}$, $\alpha_\mathrm{D}$, $\alpha_\mathrm{o}$, and $\alpha_\mathrm{v}$ are positive scalable parameters, which can be tuned for better estimator design. If the scalable parameters
\begin{equation*}
  \boldsymbol \alpha = [\alpha_\mathrm{P}\;\;\alpha_\mathrm{I}\;\;\alpha_\mathrm{D}\;\;\alpha_\mathrm{o}\;\;\alpha_\mathrm{v}]^\mathrm{T}
\end{equation*}
are known, we can directly obtain $\mathbf{S}$.

Substituting \eqref{eq:ycomponents} and \eqref{eq:S} into \eqref{eq:zk} yields
\begin{equation*}
  \mathbf{z}_k=\mathbf{z}_k^\mathrm{P}+\mathbf{z}_k^\mathrm{I}+\mathbf{z}_k^\mathrm{D}+\mathbf{z}_k^\mathrm{o}+\mathbf{z}_k^\mathrm{v}+\mathbf{b}
\end{equation*}
where
\begin{eqnarray*}
\mathbf{z}_k^\mathrm{P} &=& \alpha_\mathrm{P}\mathbf{R}_1\mathbf{c}_{k-l:k}\\
\mathbf{z}_k^\mathrm{I} &=& \alpha_\mathrm{I}\mathbf{R}_2\mathbf{c}_{k-l+s:k}^\mathrm{I}\\ \mathbf{z}_k^\mathrm{D} &=& \alpha_\mathrm{D}\mathbf{R}_3\mathbf{c}_{k-l+1:k}^\mathrm{D}\\ \mathbf{z}_k^\mathrm{o} &=& \alpha_\mathrm{o}\mathbf{R}_4\mathbf{o}_{k-l:k-1}\\
\mathbf{z}_k^\mathrm{v} &=& \alpha_\mathrm{v}\mathbf{R}_5\mathbf{v}_{k-l:k}
\end{eqnarray*}
and $\mathbf{R}=[\mathbf{R}_1\; \mathbf{R}_2\;\mathbf{R}_3\;\mathbf{R}_4\;\mathbf{R}_5]$. It is clear that the scalable parameters can adjust the influence of different features on $\mathbf{z}_k$. If all the five scalable parameters are equal like \eqref{eq:S_equal}, the entries of $\mathbf{z}_k^\mathrm{P}$ will overwhelm those of the other four components because the value of CO2 data is much larger than the data of the other features. In that case, compared with the CO2 data the other four features are insignificant.

In practice, we use the following three steps to determine the scalable parameters:
\begin{enumerate}
  \item Generate the random matrix $\mathbf{R}$ and calculate $z_\mathrm{max}^\mathrm{P^\prime}, z_\mathrm{max}^\mathrm{I^\prime}$, $z_\mathrm{max}^\mathrm{D^\prime}, z_\mathrm{max}^\mathrm{o^\prime}$, and $z_\mathrm{max}^\mathrm{v^\prime}$ from the training data. Here
      \begin{equation*}
        z_{\mathrm{max}}^{\mathrm{P}^\prime}=\mathbf{max}\{\|\mathbf{z}_k^{\mathrm{P}^\prime}\|_\infty =\|\mathbf{R}_1\mathbf{c}_{k-l:k}\|_\infty, \; k\leq N\}
      \end{equation*}
      and $\mathbf{z}_k^{\mathrm{P}^\prime}=\frac{1}{\alpha_\mathrm{P}}\mathbf{z}_k^{\mathrm{P}}$. Like $z_{\mathrm{max}}^{\mathrm{P}^\prime}$, we have the corresponding similar definitions of $z_\mathrm{max}^\mathrm{I^\prime}$, $z_\mathrm{max}^\mathrm{D^\prime}, z_\mathrm{max}^\mathrm{o^\prime}$, and $z_\mathrm{max}^\mathrm{v^\prime}$.

  \item Set $z_\mathrm{max}^\mathrm{P}, z_\mathrm{max}^\mathrm{I}, z_\mathrm{max}^\mathrm{D}$, $z_\mathrm{max}^\mathrm{o}$, and $z_\mathrm{max}^\mathrm{v}$ based on the following equation
      \begin{equation}\label{eq:zmax}
          z_\mathrm{max}^\mathrm{P}+z_\mathrm{max}^\mathrm{I}+z_\mathrm{max}^\mathrm{D}+ z_\mathrm{max}^\mathrm{o}+z_\mathrm{max}^\mathrm{v}<5
      \end{equation}
      where $z_{\mathrm{max}}^{\mathrm{P}}=\mathbf{max}\{\|\mathbf{z}_k^{\mathrm{P}}\|_\infty, \; k\leq N\}$.
  \item Evaluate the scalable parameters by $\alpha_\mathrm{s}=\frac{z_\mathrm{max}^\mathrm{s}}{z_\mathrm{max}^{\mathrm{s}^\prime}}$ where `$\mathrm{s}$' corresponds to the subscripts/superscripts `$\mathrm{P}$', `$\mathrm{I}$', `$\mathrm{D}$', `$\mathrm{o}$', `$\mathrm{v}$', respectively.
\end{enumerate}
Apparently, the five parameters in \eqref{eq:zmax} stand for the influence of the five features on the hidden neurons (i.e., $\mathbf{z}_k$ and $\mathbf{h}(\mathbf{z}_k)$). In practice, we can tune the five parameters and check the effectiveness of the occupancy regression model via cross validation to find a group of effective $\boldsymbol\alpha$. With the tuned scalable parameters $\boldsymbol\alpha$ we can obtain the corresponding scale matrix $\mathbf{S}$ from \eqref{eq:S}. It is obvious that \eqref{eq:S_equal} is a special case of \eqref{eq:S}.
On one hand the scale matrix in \eqref{eq:S}  scales $\|\mathbf{z}_k\|_\infty$ to a proper value (less than 5), on the other hand it balances the influence of different features.

Next, we determine the hidden-to-output layer connection weights. Considering the training data set $\{(\mathbf{x}_k, o_k)| k=1,2\cdots N\}$, \eqref{eq:IS-ELM} can be rewritten as the following compact form:
\begin{equation}\label{eq:compactELM}
  \mathbf{H}\boldsymbol\beta=\mathbf{O}
\end{equation}
where $\mathbf{H}$ is the hidden layer output matrix
\begin{equation*}
  \mathbf{H}=\left[
               \begin{array}{c}
                 \mathbf{h}(\mathbf{z}_1) \\
                 \vdots \\
                 \mathbf{h}(\mathbf{z}_N) \\
               \end{array}
             \right]
  =\left[
     \begin{array}{ccc}
       h(z_{11}) & \cdots & h(z_{1L}) \\
       \vdots& \vdots & \vdots \\
       h(z_{N1}) & \cdots & h(z_{NL}) \\
     \end{array}
   \right]
\end{equation*}
and $\mathbf{O}=[o_1, o_2 \cdots o_N]^\mathrm{T}$. With the designed scalable parameters, $\mathbf{H}$ can be easily calculated. Using the least-squares technique, the solution of \eqref{eq:compactELM} is given by $\boldsymbol\beta=\mathbf{H}^\dag\mathbf{O}$

To improve the generalization performance and make the solution robust, a regularization term is added into to the least-squares solution. As seen from \cite{huang2012extreme}, based on the $l_2$ norm optimization equations, the solution of output weights becomes
\begin{equation}\label{eq:beta}
  \boldsymbol\beta = \left(\gamma\mathbf{I}+\mathbf{H}^\mathrm{T}\mathbf{H}\right)^{-1}\mathbf{H}^\mathrm{T}\mathbf{O}
\end{equation}
where $\gamma$ is a regularization parameter.

$\mathbf{W}_1$ is known \emph{a priori}. With scalable parameters $\boldsymbol\alpha$ we can obtain $\mathbf{S}$. $\mathbf{R}$ and $\mathbf{b}$ are randomly generated and $\boldsymbol\beta$ can be obtained from \eqref{eq:beta}. With $\mathbf{W}_1, \mathbf{R}, \mathbf{b}$, $\mathbf{S}$,and $\boldsymbol\beta$, from \eqref{eq:IS-ELM} we can find the occupancy estimator
\begin{equation}\label{eq:Occestimator}
  o_k = f(\mathbf{x}_k)=\mathbf{h}(\mathbf{RS}\mathbf{W}_1\mathbf{x}_k+\mathbf{b})\boldsymbol\beta
\end{equation}

Like the standard ELM, in the training process the main computational cost is attributed to solving the inverse matrix in \eqref{eq:beta}. Hence, the FS-ELM retains the computational efficiency of the standard ELM.

\section{smoothed CO2 Data for occupancy estimation}

\subsection{Estimator identification from globally smoothed CO2 data}

\begin{figure}[ht]
    \centering
    \includegraphics[width=0.48\textwidth]{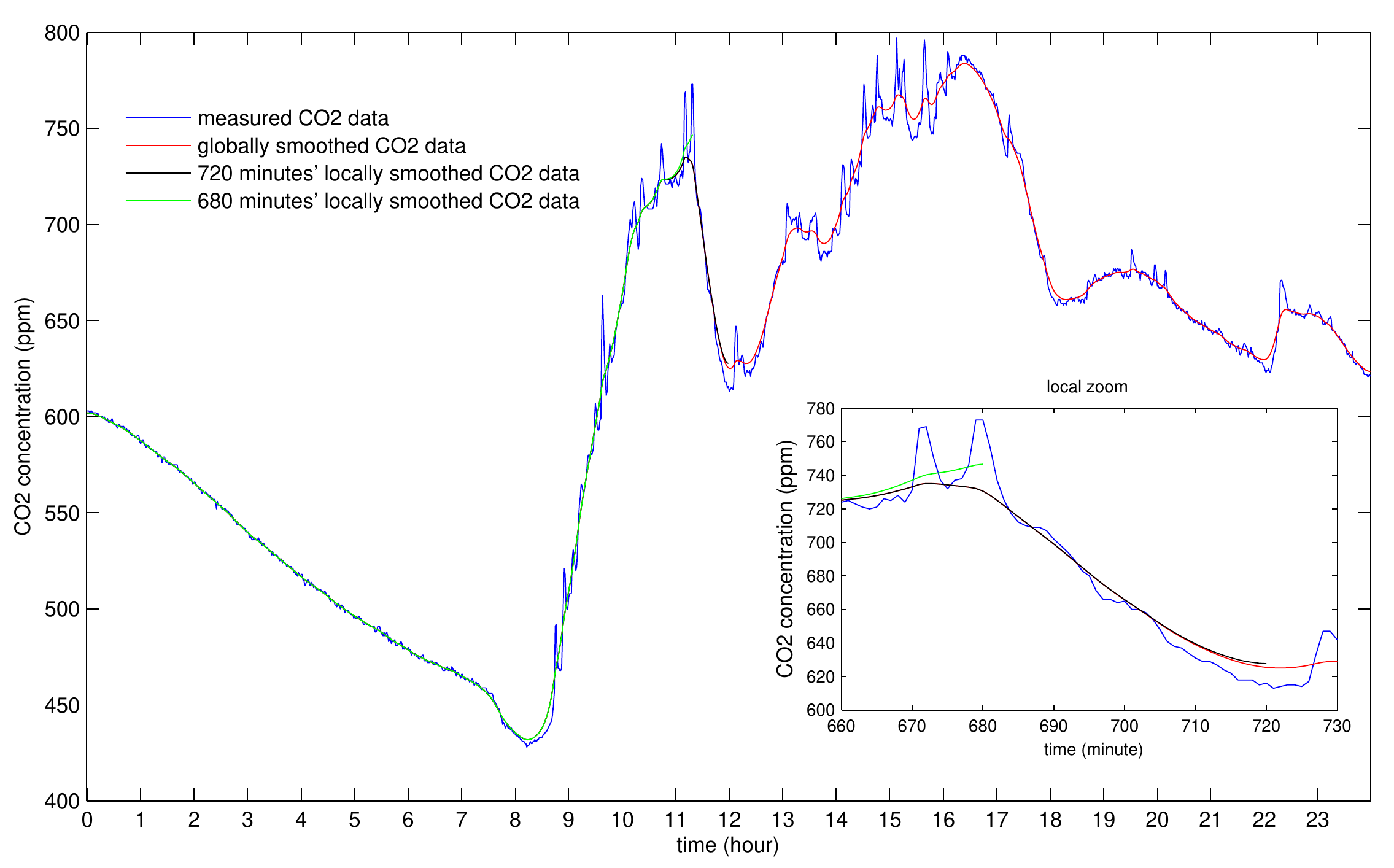}
       \caption{One day CO2 concentration of the test-bed.}
    \label{Fig:SCO2}
\end{figure}

In most of real applications, indoor CO2 concentration is non-uniform \cite{weekly2015modeling}. It is shown in Fig. \ref{Fig:SCO2} that the measured CO2 concentration suffers from serious spikes, which may be induced by the measurement noise, irregular indoor air movement and occupants' irregularly approaching the sensor. The influence of the irregular air and occupants movement on the CO2 concentration of a certain location is still an open problem. Obviously, the measured CO2 concentration with spikes cannot reflect the CO2 level of the whole indoor space, which should be used for occupancy estimation.  Therefore, to offset the negative influence of the spikes in the measured CO2 concentration, we smooth the measured data.

We denote all the measured CO2 concentration data by a vector $\mathbf{c}=[c_1, c_2,\cdots, c_N]^\mathrm{T}$. The smoothed CO2 data $\mathbf{c}_\mathrm{s}$ can be obtained by minimizing the following energy function
\begin{equation}
J(\mathbf{c}_\mathrm{s}) = \|\mathbf{c}-\mathbf{c}_\mathrm{s}\|_2^2+\lambda \|\nabla \mathbf{c}_\mathrm{s}\|_2^2 \label{eq:costfunction}
\end{equation}
where $\nabla \mathbf{c}_\mathrm{s}$ represents the gradient of $\mathbf{c}_\mathrm{s}$,
\begin{equation*}
  \nabla=\left[
           \begin{array}{cccc}
             -1 & 1 &  &  \\
                & \ddots & \ddots &  \\
              &  & -1 & 1\\
           \end{array}
         \right]\in \mathbb{R}^{N-1\times N},
\end{equation*}
and $\lambda$ is a weighting factor which controls the balance between the two terms in \eqref{eq:costfunction}. Increasing $\lambda$ results in more smoothing of the output $\mathbf{c}_\mathrm{s}$.

Taking the derivative of $J(\mathbf{c}_\mathrm{s})$ in \eqref{eq:costfunction} with respect to $\mathbf{c}_\mathrm{s}$ and setting it to zero yields
\begin{equation}\label{eq:smoothfunction}
  (\mathbf{I}+\lambda\Delta)\mathbf{c}_\mathrm{s}=\mathbf{c}
\end{equation}
where $\Delta=\nabla^\mathrm{T}\nabla$ is a Laplacian matrix, and $\mathbf{I}+\lambda\Delta$ is a tridiagonal matrix. Hence \eqref{eq:smoothfunction} can be solved via the Thomas algorithm in $O(N)$ operations \cite{datta2010numerical}. From \eqref{eq:smoothfunction}, we can easily obtain the smoothed CO2 concentration
\begin{equation}\label{eq:smoothedCO2}
  \mathbf{c}_\mathrm{s}=(\mathbf{I}+\lambda\Delta)^{-1}\mathbf{c}
\end{equation}
The smoothed CO2 concentration is shown in Fig. \ref{Fig:SCO2}. Such a smoother is commonly used for image/signal edge-preserving smoothing \cite{min2014fast, cho2016effective}. 

Using the smoothed CO2 data we can train a new occupancy estimator via the FS-ELM introduced in Section III, which we denote by
\begin{equation}\label{eq:SmoothedOccEst}
  o_k=f_\mathrm{s}(\mathbf{x}_k^\mathrm{gs})=f_\mathrm{s}(\mathbf{c}_{k-l:k}^\mathrm{gs}, \mathbf{o}_{k-l:k-1}, \mathbf{v}_{k-l:k})
\end{equation}
where the superscript `gs' means the `global smoothing', which will be discussed later. The occupancy estimator identified from the smoothed CO2 data outperforms the estimator identified directly from the raw measured CO2 data, which is shown in Section \ref{Sec:experiment} via the experiment.

\subsection{Occupancy estimation from locally smoothed CO2 data}
When using the identified occupancy estimator, $\mathbf{o}_{k-l:k-1}$, i.e. the exact number of occupants in the time horizon $[t_{k-l}, t_{k-1}]$, is generally unknown. In practice, we can use the estimated occupancy level $\hat{\mathbf{o}}_{k-l:k-1}$ instead of $\mathbf{o}_{k-l:k-1}$. Therefore, the practically used estimator \eqref{eq:Occestimator} becomes
\begin{equation}\label{eq:HatOELM}
    \hat{o}_k = f(\hat{\mathbf{x}}_k)=  f(\mathbf{c}_{k-l:k}, \hat{\mathbf{o}}_{k-l:k-1}, \mathbf{v}_{k-l:k})
\end{equation}

Similarly, the occupancy estimator \eqref{eq:SmoothedOccEst} trained by the globally smoothed CO2 data becomes
\begin{equation}\label{eq:HatOSmooth}
  \hat{o}_k^\mathrm{gs} = f_\mathrm{s}(\hat{\mathbf{x}}_k^\mathrm{gs})=  f_\mathrm{s}(\mathbf{c}_{k-l:k}^\mathrm{gs}, \hat{\mathbf{o}}_{k-l:k-1}^\mathrm{gs}, \mathbf{v}_{k-l:k})
\end{equation}
where $\mathbf{c}_{k-l:k}^\mathrm{gs}$ consists of the $(k-l)$-th to $k$-th entries of the smoothed CO2 data
\begin{equation}\label{eq:SmoothedCO2Testing}
\mathbf{c}^\mathrm{gs}=(\mathbf{I}+\lambda\Delta_M)^{-1}\mathbf{c}^\mathbf{u}
\end{equation}
Here, $M$ implies the dimension of the the Laplacian matrix $\Delta_M$, and $\mathbf{c}^\mathrm{u}\in\mathbb{R}^M$ is measured CO2 data for an entire day, which is real-time unknown when using the estimator \eqref{eq:HatOSmooth}.

Equation \eqref{eq:smoothedCO2} and \eqref{eq:SmoothedCO2Testing} show that we need global information of the measured CO2 data (i.e. all CO2 data used for training or a whole day of CO2 data for testing/using) to find the smoothed CO2 concentration. In the process of training the occupancy estimator, all measured CO2 data are \emph{a prior} known and we can easily obtained the smoothed CO2 data from \eqref{eq:smoothedCO2}. When using the occupancy estimator, however, the entire day's CO2 data is real-time unknown. Only the current and past CO2 data are measured and future data are unknown. In real applications, only the locally smoothed CO2 data can be real-time obtained.

In practice, $\mathbf{c}^\mathrm{u}$ and $\mathbf{c}^\mathrm{gs}$ in \eqref{eq:SmoothedCO2Testing} and $\mathbf{c}_{k-l:k}^\mathrm{gs}$ in \eqref{eq:HatOSmooth} is real-time unavailable and hence we cannot use \eqref{eq:HatOSmooth}. An intuitive idea is to replace $\mathbf{c}_{k-l:k}^\mathrm{gs}$ by the locally smoothed CO2 data $\mathbf{c}_{k-l:k}^\mathrm{ls}$, and the estimator is described as
\begin{equation}\label{eq:HatOLocalSmooth}
  \hat{o}_k^\mathrm{ls} = f_\mathrm{s}(\hat{\mathbf{x}}_k^\mathrm{ls})=  f_\mathrm{s}(\mathbf{c}_{k-l:k}^\mathrm{ls}, \hat{\mathbf{o}}_{k-l:k-1}^\mathrm{ls}, \mathbf{v}_{k-l:k})
\end{equation}
where $\mathbf{c}_{k-l:k}^\mathrm{ls}$ consists of the $(k-l)$-th to the $k$-th entries of the local smoothed CO2 data
\begin{equation}\label{eq:LocalSmoothedCO2Testing}
\mathbf{c}_{1:k}^\mathrm{ls}=(\mathbf{I}+\lambda\Delta_k)^{-1}\mathbf{c}_{1:k}
\end{equation}
The black line in Fig. \ref{Fig:SCO2} is the locally smoothed 720 minutes CO2 data, which almost coincides with the globally smoothed CO2 data. The green line shows that the locally smoothed 680 minutes CO2 data are different with the globally smoothed data in the region near the end point, i.e. 11:20 am. In practice, the end point exactly corresponds to the current time instant. It is not difficult to see from Fig. \ref{Fig:SCO2} that the locally smoothed data almost coincides with the globally smoothed data except the region near the end point, and the difference in this region is small compared with the magnitude of the spikes.

However, directly using the estimator in \eqref{eq:HatOLocalSmooth} suffers from two problems. Firstly, the regression model $f_\mathrm{s}(\cdot)$ is identified from the globally smoothed CO2 data but the real-time locally smoothed CO2 data is slightly different with the globally smoothed data. Secondly, $\hat{\mathbf{o}}_{k-l:k-1}^\mathrm{ls}$ is a part of the input data to estimate $\hat{\mathbf{o}}_{k}^\mathrm{ls}$. If the estimated occupancy level $\hat{\mathbf{o}}_{k-l:k-1}^\mathrm{ls}$ is not accurate due to the difference between the locally and globally smoothed CO2 data, the error will be transmitted to $\hat{o}_k^{\mathrm{ls}}$. The estimation error is easily accumulated, and the accumulated error makes the results of \eqref{eq:HatOLocalSmooth} meaningless when $k$ is large.

To remove the accumulated error, at the time instant $t_k$, we carry out the following three steps:
\begin{enumerate}
  \item Evaluate the locally smoothed CO2 data $\mathbf{c}_{1:k}^\mathrm{ls}$ from \eqref{eq:LocalSmoothedCO2Testing}.
  \item Evaluate the occupancy level $\hat{\mathbf{o}}_{l+1:k|k}^\mathrm{s}$. For all $l+1\leq i\leq k$,
          \begin{equation}\label{eq:HatOLocalSmoothTesting}
          \hat{o}_{i|k}^\mathrm{s} =   f_\mathrm{s}(\mathbf{c}_{i-l:i|k}^\mathrm{s}, \hat{\mathbf{o}}_{i-l:i-1|k}^\mathrm{s}, \mathbf{v}_{i-l:i})
        \end{equation}
        where $\mathbf{c}_{i-l:i|k}^\mathrm{s}$ consists of the $(i-l)$-th to the $i$-th entries of $\mathbf{c}_{1:k}^\mathrm{ls}$ obtained from step 1).
  \item The estimated occupancy level at $t_k$ is $\hat{o}_{k|k}^\mathrm{s}$.
\end{enumerate}
At each sampled time instant we need to estimate all the past occupancy levels. But in the process of estimating $\hat{o}_{k|k}^\mathrm{s}$ we do not use $\{\hat{o}_{k-l|k-l}^\mathrm{s},...,\hat{o}_{k-1|k-1}^\mathrm{s}\}$. Therefore, the estimation error cannot be transmitted and we can avoid the accumulation of the estimation error.

In addition, when using the occupancy estimators, the initial state of the occupancy level is required. We can reset the estimator at every midnight and set the initial state as $\hat{\mathbf{o}}_{0:l-1}=\mathbf{0}\in \mathbb{R}^{l}$.

\begin{figure}[ht]
    \centering
    \includegraphics[width=0.45\textwidth]{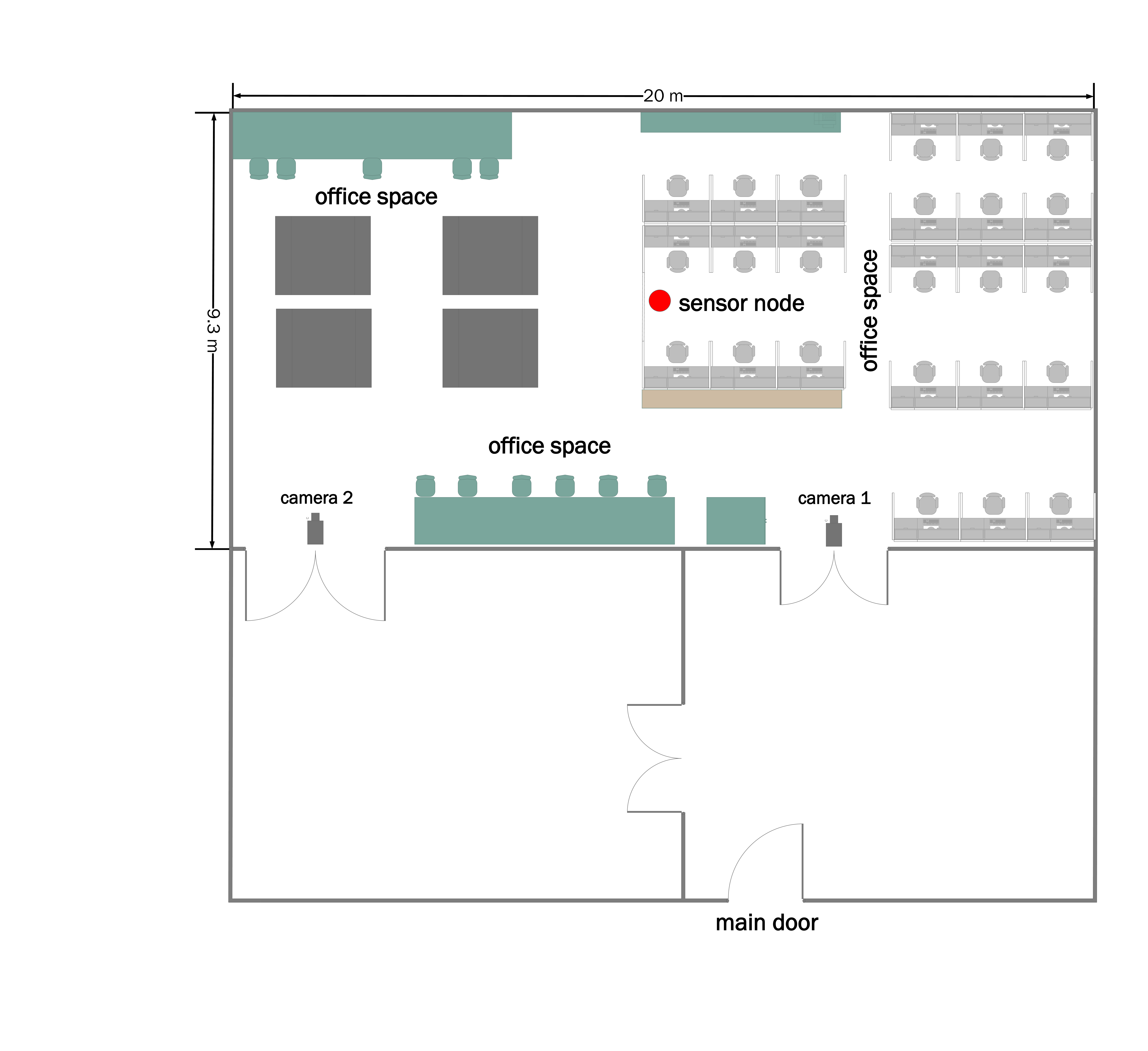}
       \caption{The layout of the test-bed.}
    \label{Fig:Room}
\end{figure}

\section{Experiment verification}\label{Sec:experiment}

\subsection{Statement of the experiment}
We tested the proposed occupancy estimators in the Process Instrumentation Lab in Nanyang Technological University, Singapore. The layout of the test-bed is shown in Fig. \ref{Fig:Room}. The laboratory contains a 9.3m$\times$ 20m office space with 24 office cubic for research students and staff, and 11 open seats for final year undergraduate students. The five open seats in the corner are rarely occupied.

A CL11 sensor from Rotronic was used for CO2 measurement. The resolution of the sensor is 1 ppm. The sampling time was 1 minute. The location of the sensor is in the region around the center of the office space. We select a location in this region where the sensor is easily fixed. The optimal sensor placement for occupancy estimation is still an open problem and out of the scope of this paper.

The laboratory was installed both Variable Air Volume (VAV) and Active Chilled Beam (ACB) systems for air conditioning. The VAV system is always turn on while the ACB system is unscheduled on. The office space is mechanically ventilated using an Air Handling Unit (AHU) which delivers a constant supply of fresh air. Everyday the ventilation system is turn on at 7:30 AM and turn off at 10:00 PM.
In this work, we set the venting level as a binary state, i.e. using 1 and 0 represent the ON and OFF state, respectively.

\subsection{The ground truth acquisition}
As shown in Fig. \ref{Fig:Room}, we set two cameras to record the entrance and exit of the two doors of the office space. Since the sampling time of the CO2 sensor is one minute, we need to firstly obtain the number of entrance and exit of the two doors for every minute from the videos.
With all historical data of the number of entrances and exits, we can easily calculate the number of occupants in the office space at any time.


To obtain the ground truth, i.e., the exact number of the occupants in the office space, from the videos of the two cameras, we do the following:
\begin{enumerate}
  \item First, we found all the minutes (i.e. the time interval of one minute) in which occupants appeared in one of the videos using computer vision algorithm (histogram of oriented gradient)\cite{dalal2006human};
  \item Then, we manually counted the exact number of entrance and exit of the two doors for all the time intervals in step 1);
  \item Finally, we calculated the occupancy level in the office space based on the historical entrance and exit data of the two doors obtained from step 2).
\end{enumerate}
In this way, we have collected 30 weekdays of occupancy data of the office space. All the calculated occupancy levels of the 30 days are zero after 11:30 pm, which implies the collected occupancy data is almost exact. We used the 30 days of data as the ground truth.

\subsection{The performance indexes}

To assess the performance of the proposed occupancy estimator, we consider the five performance indices:
\begin{itemize}
  \item \emph{root mean squared error} (RMSE), indicating the mean magnitude of the estimation error as defined in \eqref{eq:RMSE};
  \item\emph{ false positive rate} (FPR), representing the rate that the room is estimated to be occupied while it is not, as defined in \eqref{eq:FPR};
  \item \emph{false negative rate} (FNR), representing the rate that the room is estimated to be empty while it is not, as defined in \eqref{eq:FNR};
  \item \emph{false detection rate} (FDR), representing the rate of false detecting whether the room is occupied or not, as defined in \eqref{eq:FDR};
  \item \emph{$x$-tolerance accuracy}, reporting the percentage that the occupancy estimator can provide the estimates whose errors are less than $x$, as defined in \eqref{eq:xaccuracy}.
\end{itemize}

Next we present the definitions of the five performance indices. The RMSE is defined by
\begin{equation}\label{eq:RMSE}
  \mathrm{RMSE}(\hat{\mathbf{o}}):= \sqrt{\frac{\|\mathbf{o}-\hat{\mathbf{o}}\|_2^2}{M}}
\end{equation}
where $\hat{\mathbf{o}}\in\mathbb{R}^M$ is the estimated occupancy level in $M$ sampling time, and $\mathbf{o}$ represents the real occupancy level.

FPR, FNR and FDR are used to assess the performance in occupancy detection. FPR and FNR have been presented in \cite{ebadat2015regularized}. To define the three indices, we introduce the following indicator function
\begin{equation*}\label{eq:l0function}
  \mathds{1}(o_k):= \begin{cases}
                      1, & \mbox{if } o_k>0 \\
                      0, & \mbox{otherwise}.
                    \end{cases},\quad
  \mathds{1}(\mathbf{o}_k):= \left[
                               \begin{array}{c}
                                 \mathds{1}(o_1) \\
                                 \vdots \\
                                 \mathds{1}(o_M) \\
                               \end{array}
                             \right]
\end{equation*}
and the time set
\begin{equation*}
  \mathcal{N}_\theta:=\{k|\mathds{1}(o_k)=\theta, k\in \mathbb{N}^+, \mbox{and } k\in[1,M]\}
\end{equation*}
It is clear that $\mathcal{N}_0$ consists of all the time indices for which the office space was not occupied, while $\mathcal{N}_1$ consists of the time indices for which the office space was occupied. Therefore, we define
\begin{eqnarray}
  \mathrm{FPR}(\hat{\mathbf{o}}) &=& \frac{1}{|\mathcal{N}_0|}\underset{k\in \mathcal{N}_0}{\sum}\mathds{1}(\hat{o}_k)\label{eq:FPR} \\
  \mathrm{FNR}(\hat{\mathbf{o}}) &=& 1-\frac{1}{|\mathcal{N}_1|}\underset{k\in \mathcal{N}_1
  }{\sum}\mathds{1}(\hat{o}_k)\label{eq:FNR}\\
  \mathrm{FDR}(\hat{\mathbf{o}}) &=& \frac{1}{M}\left[\underset{k\in \mathcal{N}_0}{\sum}\mathds{1}(\hat{o}_k)+ |\mathcal{N}_1|-\underset{k\in \mathcal{N}_1 }{\sum}\mathds{1}(\hat{o}_k)\right]\label{eq:FDR}
\end{eqnarray}

Most of the current research work in terms of occupancy estimation used the \emph{accuracy} as one performance index, which represents the percentage that the estimator returns the correct value, and is defined by
\begin{equation*}\label{eq:accuracy}
  \mathrm{Acc}(\hat{\mathbf{o}}):=\frac{M-\sum_{k=1}^M\mathds{1}(|o_k-\hat{o}_k|)}{M}
\end{equation*}
The notion of \emph{accuracy} is a proper criterion if the indoor occupants is only a few (e.g., less than 4 in \cite{ebadat2015regularized, dong2010information, lam2009occupancy}). But if the number of indoor occupants is large, $\mathrm{Acc}(\hat{\mathbf{o}})$, cannot well reflect the performance of the estimator. For example, if $o_k=20$, the estimate $\hat{o}_k=21$ is totally wrong when using the \emph{accuracy }as an assessment criterion. But in many real applications like indoor air-conditioning or lighting control systems, 20 or 21 occupants probably have no practical difference for taking certain operation decision.

To well assess the performance of general occupancy estimators, we introduce the following $x$-\emph{tolerance accuracy}
\begin{equation}\label{eq:xaccuracy}
  \tau(\hat{\mathbf{o}},x):=\frac{\sum_{k=1}^M\mathds{X}(|o_k-\hat{o}_k|,x)}{M}
\end{equation}
where
\begin{equation*}
  \mathds{X}(|o_k-\hat{o}_k|,x)=\begin{cases}
                                  1, & \mbox{if } |o_k-\hat{o}_k|\leq x \\
                                  0, & \mbox{otherwise}.
                                \end{cases}
\end{equation*}
We can easily find that the \emph{accuracy} is a special case of the $x$-\emph{tolerance accuracy}, i.e.
$\tau(\hat{\mathbf{o}},0)=\mathrm{Acc}(\hat{\mathbf{o}})$. The tolerance $x$ should be properly selected based on the specific applications.

\begin{figure*}[ht]
    \centering
    \includegraphics[width=0.95\textwidth]{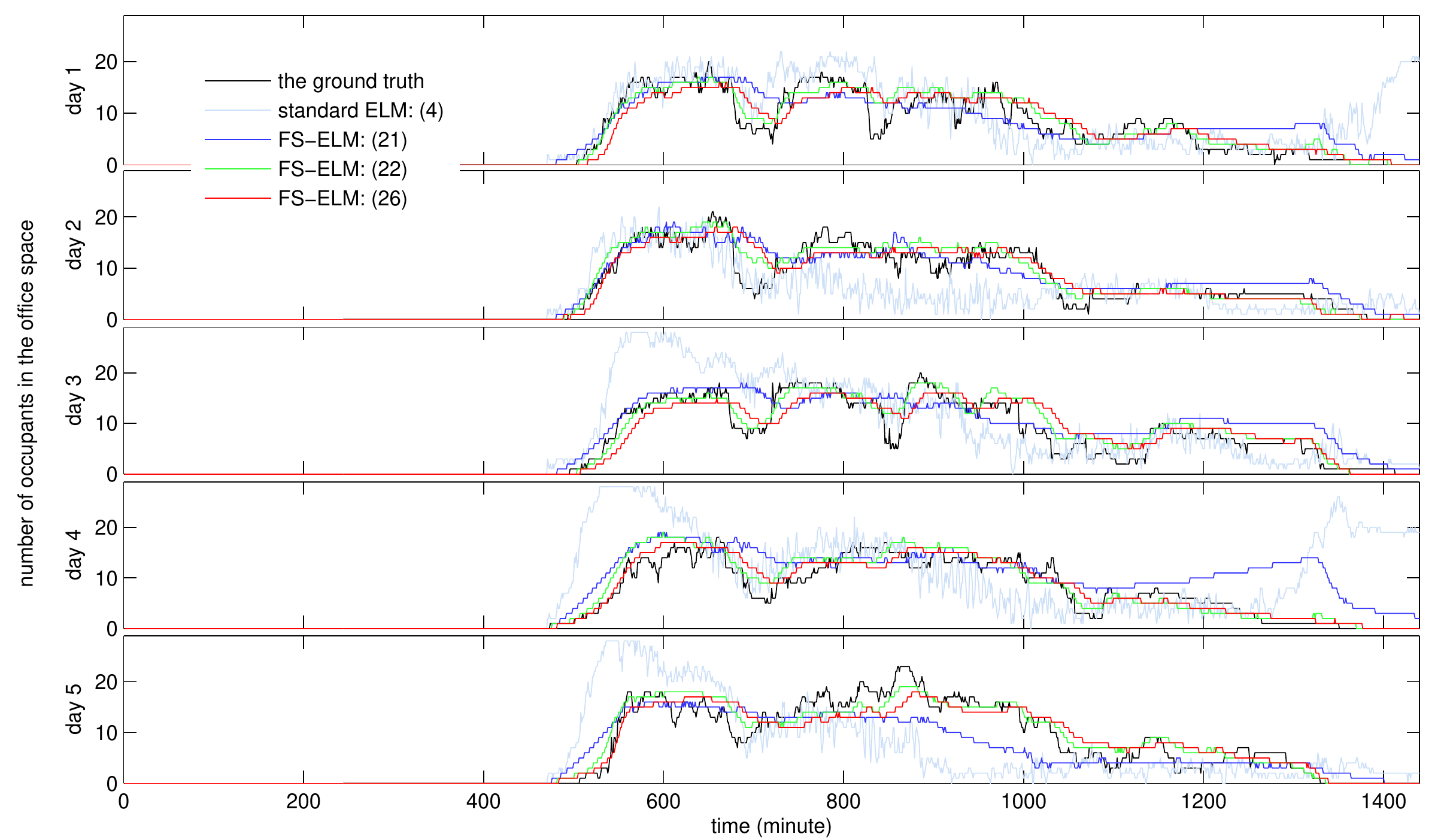}
       \caption{The occupancy estimation results of the five testing days. The x-axes is the time. Noted that one day is with 1440 minutes. In both training and testing processes, we used the measured CO2 data for the estimators \eqref{eq:ELMoutput} and \eqref{eq:HatOELM}. We used the globally smoothed CO2 data to train the estimator \eqref{eq:SmoothedOccEst}. The green lines show the results of \eqref{eq:HatOSmooth} which used the globally smoothed CO2 data for testing, while the red lines show the results $\hat{o}_{k|k}^\mathrm{s}$ for $k=1,\ldots M$ in \eqref{eq:HatOLocalSmoothTesting}, which used the locally smoothed CO2 data for testing. Noted that we can obtain the globally smoothed data for testing. But in real applications we cannot obtain the real-time globally smoothed data when using the estimator.}
    \label{Fig:Results}
\end{figure*}

\subsection{The parameters setting and results}

We collected 30 days of data of the office space. 25 days of data were used to train the occupancy estimator, and the other five days data were used for testing.
For both ELM and FS-ELM, all entries of the random matrices $\mathbf{W}$ and $\mathbf{R}$ were set to follow uniform distribution in between -1 and 1, i.e. $w_{ij}\sim\mathcal{U}(-1,1)$ and $r_{ij}\sim \mathcal{U}(-1,1)$. All entries of the bias vector was set by $b_i\sim \mathcal{U}(-0.1,0.1)$. All the required parameters in the training process are shown in Table \ref{Table:parameters}.

\begin{table}[htb]
  \centering
  \caption{All required parameters used in training the occupancy estimators via ELM and FS-ELM}\label{Table:parameters}
  \begin{tabular}{cccccc}
  \hline\hline
  Estimator &$l$ & $s$ & $L$ & $\gamma$ & $\lambda$  \\ \hline
  (4) (16) \& (20) & 30 & 10 & 1000 & 0.001 & 50\\ \hline\hline
  & $\alpha_\mathrm{P}$ & $\alpha_\mathrm{I}$ & $\alpha_\mathrm{D}$ & $\alpha_\mathrm{o}$ & $\alpha_\mathrm{v}$ \\ \hline
  (16) & 0.0001 & 0.0012 & 0.005 & 0.004 & 0.008\\
  (20) & 0.0001 & 0.0025 & 0.02  & 0.004 & 0.008\\ \hline \hline
\end{tabular}
\end{table}

We determined the scalable parameters $\boldsymbol\alpha$  based on the method provided in Section \ref{SubSec:training}. We simply set the parameters $z_\mathrm{max}^\mathrm{P}, z_\mathrm{max}^\mathrm{I}, z_\mathrm{max}^\mathrm{D}$, $z_\mathrm{max}^\mathrm{o}$, and $z_\mathrm{max}^\mathrm{v}$, which stand for the influence of the five features on the hidden neurons, as 1, 1, 1, 1, and 0.1, respectively.
If we set a larger $z_\mathrm{max}^\mathrm{v}$ such as 0.5, the solutions of the FS-ELM may have a jump after 10:00 PM (the ventilation system is turned off), like the results of the standard ELM, which will be mentioned later. If the random matrix $\mathbf{R}$ changed, the obtained scalable parameters would slightly change. The scalable parameters in  Table \ref{Table:parameters} are the approximated solutions based on one random matrix $\mathbf{R}$. We have checked that if we regenerated $\mathbf{R}$ whose entries $r_{ij}\sim \mathcal{U}(-1,1)$, the parameters $z_\mathrm{max}^\mathrm{P}, z_\mathrm{max}^\mathrm{I}, z_\mathrm{max}^\mathrm{D}$, $z_\mathrm{max}^\mathrm{o}$, and $z_\mathrm{max}^\mathrm{v}$ were still around 1, 1, 1, 1, and 0.1, respectively. We remark that all the parameters in Table \ref{Table:parameters} can still be further tuned for better performance of the estimator.

The results of the five days of occupancy estimation is shown in Fig. \ref{Fig:Results}. In the training process of ELM/FS-ELM, we generated 100 random matrices ($\mathbf{W}$ for ELM and $\mathbf{R}$ for FS-ELM), and selected the one leading to minimum training error.
In the 25 days data, the maximum number of occupants is 28; therefore, in testing the occupancy estimators if the estimated occupancy level is more than 28 we modified it to be 28. In addition, if the estimated occupancy level is negative, we set it as zero.

As shown in Fig. \ref{Fig:Results}, the results of the standard ELM suffer from serious fluctuations, which implies that the estimator \eqref{eq:ELMoutput} is sensitive to the noise/fluctuation of the CO2 data. In addition, for day 1 and day 4, after 10:00 PM the estimated occupancy level from the standard ELM jumps to  quite a high level while the real occupancy level is almost zero. At 10:00 PM the ventilation system was turned off and the CO2 level would have a jump as can be seen in Fig. \ref{Fig:SCO2}. We have analyzed in Section \ref{Sec:Identification} that the value of CO2 data is much larger than the venting level (0 or 1), which leads to the overwhelming influence of CO2 data over that of the venting level in the estimator \eqref{eq:ELMoutput}.
However, the proposed FS-ELM can overcome this drawback, as can be seen in the blue lines in Fig. \ref{Fig:Results} having a great improvement.

Globally smoothing the CO2 data can further improve the performance of FS-ELM. The estimator \eqref{eq:HatOSmooth} (green lines in Fig. \ref{Fig:Results}) provides the best result amongst the four estimators, which used the globally smoothed CO2 data for both training and testing. In real applications, however, we cannot obtain the real-time globally smoothed data when using the occupancy estimator. Therefore, we proposed the way using locally smoothed data for occupancy estimation. The results (i.e. red lines in Fig. \ref{Fig:Results}) are quite similar with those of using globally smoothed data for testing.

\begin{table}[htb]
  \centering
  \caption{The performance indices of the estimated five days occupancy level}\label{Table:per-indexes}
\begin{tabular}{lcccc}
  \hline\hline
  Estimator          & RMSE &FDR & FPR    & FNR \\ \hline
  standard ELM: (4)  & 5.9575 & 0.0695 & 0.1713 & 0.0034 \\
  FS-ELM: (21)       & 3.2307 & 0.0569 & 0.1438 & 0.0005 \\
  FS-ELM: (22)       & 1.7726 & 0.0182 & 0.0268 & 0.0126 \\
  FS-ELM: (26) & 2.1345 & 0.0278 & 0.0441 & 0.0172 \\
  \hline\hline
\end{tabular}
\end{table}

The performance indices (i.e. RMSE, FDR, FPR, and FNR) of the five days' occupancy estimation are shown in Table \ref{Table:per-indexes}. Table \ref{Table:per-indexes} shows that the FS-ELM has a great improvement in terms of the RMSE when compared with the standard ELM. But the improvement in terms of the FDR is not very significant. However, smoothing the CO2 data can greatly improve the FDR.  Therefore, we conclude that the estimator identified by the FS-ELM can better track the curve of the real occupancy level compared with the standard ELM, while the fluctuation/noise of the CO2 data has a large influence in detecting whether the room is occupied. In addition, the results with respect to FPR and FNR indicate that all the estimators are more likely to make a false detection when the room is empty compared with when the room is occupied.

\begin{figure}[ht]
    \centering
    \includegraphics[width=0.45\textwidth]{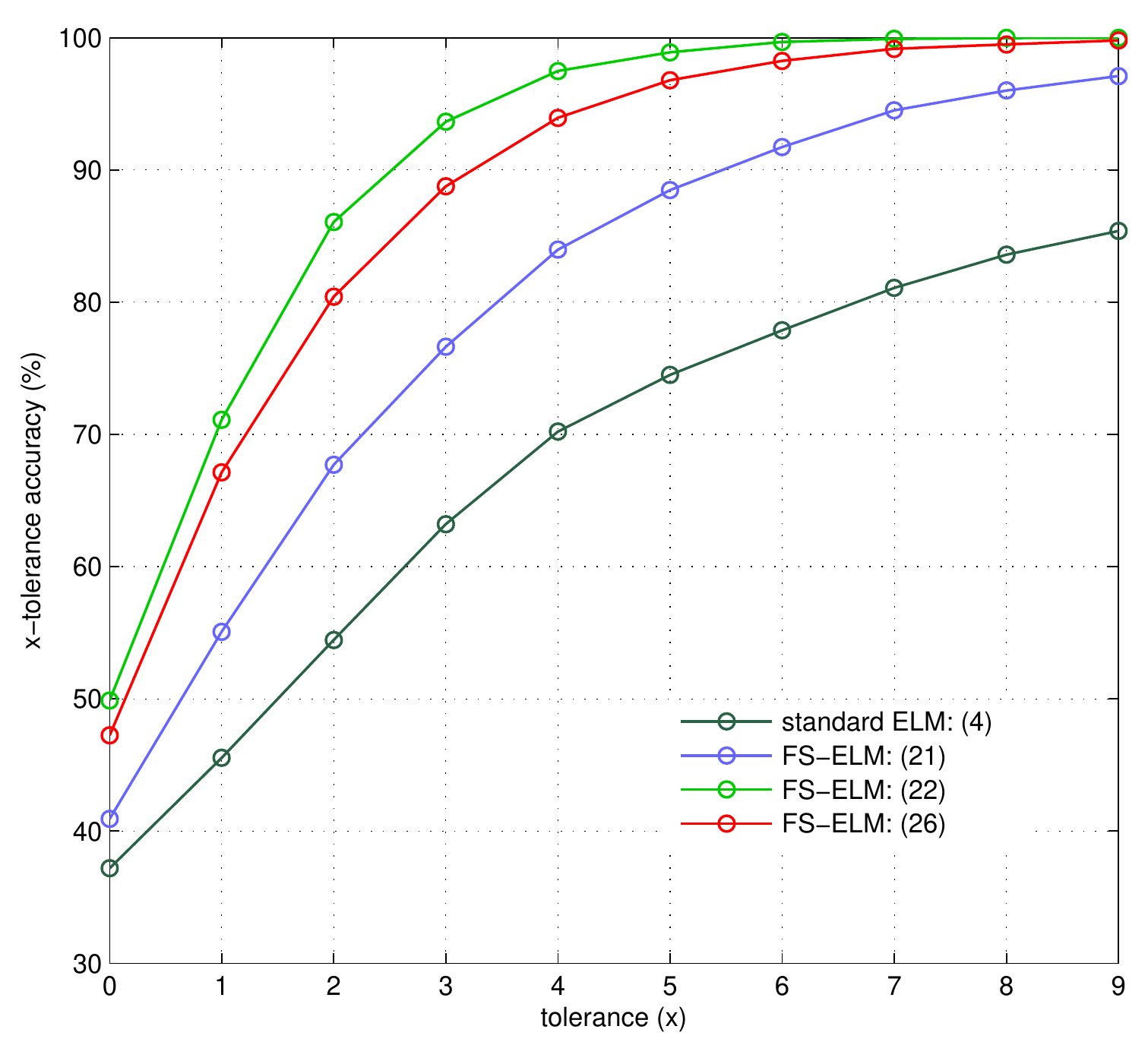}
       \caption{The $x$-\emph{tolerance accuracy} of the estimated five days' occupancy level.}
    \label{Fig:xtolAcc}
\end{figure}

The $x$-\emph{tolerance accuracy} of the results is shown in Fig. \ref{Fig:xtolAcc}. For all the four estimators, the 0-\emph{tolerance accuracy}, i.e. the \emph{accuracy}, is no more than $50\%$, which is mainly contributed by the correct occupancy detection when the room is empty, i.e. early morning and around midnight. In the day time, to estimate the exact number of indoor occupants is still a great challenge. However, for such a large office with a maximum of 35 occupants, three to four mis-estimated occupants has insignificant influence on decision-making of air-conditioning and lighting systems. We can see from Fig. \ref{Fig:xtolAcc} that three and four-tolerance accuracy is up to $89\%$ and $94\%$, respectively.

\section{Conclusions}
Indoor occupancy estimation from environmental parameters is an interesting but challenging problem, especially when the number of indoor occupants is large, such as a few tens. The proposed occupancy estimator is a discrete-time  dynamic model. Specifically, the occupancy level is a function of the CO2 data, the venting level, and the past occupancy level in a moving time horizon.

We provide a variation of the standard ELM, i.e. SF-ELM, to identify the occupancy estimator. The FS-ELM has two main difference with the standard ELM: 1) the FS-ELM has one more layer, i.e. the feature layer, which can be viewed as a preprocessing of the input layer data; and 2) the random hidden layer weight matrix is scaled based on the input layer data. The proposed FS-ELM retains the computational efficiency of the standard ELM. It greatly improves the performance of the standard ELM because 1) the input of the \emph{ELM feature mapping} (i.e. the input of the activation function) is properly scaled, which can guarantee that the SF-ELM can distinguish the difference of all the input layer data; and 2) the scaled parameters can balance the influence of difference types of feature/input layer data.

Pre-smoothing the CO2 data can further improve the results of occupancy estimation. In real applications, the globally smoothed CO2 data is real-time unavailable but the locally smoothed data is real-time available. We provide one way to remove the accumulated error deduced by the difference between global and local smoothing when we replace the globally smoothed data by locally smoothed data. 

The experiment indicates that the FS-ELM can greatly improve the performance in terms of the RMSE, while CO2 data pre-smoothing can greatly enhance the occupancy detection.

\bibliographystyle{IEEEtran}

\end{document}